\documentclass{elsart}

\usepackage{graphicx}%

\begin{document}
\begin{frontmatter}

\title{Symmetry energy from elliptic flow\\
       in $^{197}$Au + $^{197}$Au}

\author[a]{P.~Russotto},
\author[b]{P.Z.~Wu},
\author[c,d]{M.~Zoric},
\author[b]{M.~Chartier},
\author[c]{Y.~Leifels},
\author[e]{R.C.~Lemmon},
\author[f]{Q.~Li},
\author[c,g]{J.~{\L}ukasik},
\author[h]{A.~Pagano},
\author[g]{P.~Paw{\l}owski},
\author[c]{W.~Trautmann}

\thanks{corresponding author: w.trautmann@gsi.de}

\address[a]{INFN-LNS and Universit\`{a} di Catania, I-95123 Catania, Italy}
\address[b]{University of Liverpool, Physics Department, Liverpool L69 7ZE, United Kingdom}
\address[c]{GSI Helmholtzzentrum f\"{u}r Schwerionenforschung GmbH, D-64291 Darmstadt, Germany}
\address[d]{Ruder Bo\u{s}kovi\'{c} Institute. HR-10002 Zagreb, Croatia}
\address[e]{STFC Daresbury Laboratory, Warrington WA4 4AD, United Kingdom}
\address[f]{School of Science, Huzhou Teachers College, Huzhou 313000, China}
\address[g]{IFJ-PAN, Pl-31342 Krak\'ow, Poland}
\address[h]{INFN-Sezione di Catania, I-95123 Catania, Italy}

\begin{abstract} 

The elliptic-flow ratio of neutrons with respect to protons or light complex particles in
reactions of neutron-rich systems at relativistic energies is proposed as an 
observable sensitive to the strength of the symmetry term in the equation of state 
at supra-normal densities. The results obtained from the existing 
FOPI/LAND data for $^{197}$Au + $^{197}$Au collisions at 400 MeV/nucleon in comparison 
with the UrQMD model favor a moderately soft symmetry term with a density 
dependence of the potential term proportional to $(\rho/\rho_0)^{\gamma}$ 
with $\gamma = 0.9 \pm 0.4$.

\end{abstract}


\end{frontmatter}

Recent advances in the observation and modelling of astrophysical phenomena have renewed 
the interest in the nuclear equation of state (EOS) and, in particular, in its dependence 
on density and on asymmetry, i.e., on the relative neutron-to-proton 
abundance~\cite{fuchs06,klaehn06,lipr08}. 
Supernova simulations or neutron star models require inputs for the nuclear EOS 
at extreme values of these parameters~\cite{lattprak,botv04}. The dependence on asymmetry
is given by the symmetry term whose evolution with density is not only of interest for astrophysics 
but also for nuclear physics. The thickness of the neutron skin of heavy nuclei, e.g., 
reflects the differential pressure exerted on neutrons in the core~\cite{horo01}, and 
the strength of the three-body force, an important ingredient in nuclear structure 
calculations~\cite{wiringa02}, represents one of the major uncertainties in modelling the 
EOS at high density~\cite{fuchs06,xuli10}.

Considerable efforts are, therefore, underway in using heavy-ion reactions for extracting 
experimental information on the symmetry energy which is the difference between the energy 
of neutron matter and of symmetric matter. 
While fairly consistent constraints 
for the symmetry energy near normal nuclear matter density have been deduced from recent 
data~\cite{lipr08,klimk07,cent09,tsang09}, much more work is still needed to probe its high-density 
behavior. The predictions of microscopic models diverge widely there~\cite{lipr08,dyntherm}, 
and the interpretation of obvious observables 
turns out to be more difficult than perhaps anticipated.
The ratios of K$^+$/K$^0$ production 
were found to be less sensitive to the symmetry energy when 
actual collisions were modelled rather than matter under equilibrium  conditions~\cite{xlopez07}. 
Variations of the $\pi^-/\pi^+$ ratio of up to 20\% for soft versus stiff parameterizations
are expected but the competition of mean-field and collision effects appears to be very delicate. 
Different predictions leading to practically opposite conclusions were obtained from the 
comparison of the FOPI data for $^{197}$Au + $^{197}$Au collisions~\cite{reis07} 
with different transport models~\cite{ferini06,xiao09,feng10}. 

It seems essential, in this situation, to enlarge the experimental basis of suitable probes 
and to extend it from isotopic yield ratios to the isospin dependence of dynamical observables. 
Densities of up to 2 or 3 times the saturation density may be reached within a short time
scale ($\approx 20$~fm/c) in the central zone 
of heavy-ion collisions in the present range of relativistic energies of up to
$\approx 1$~GeV/nucleon~\cite{li_npa02}. 
The resulting pressure produces a collective outward motion of the compressed 
material whose strength will be influenced by the symmetry energy in asymmetric 
systems~\cite{dani02}.
Flow observables have been proposed by several groups as probes for the equation of state
at high density~\cite{li02,greco03}, among them the so-called differential neutron-proton 
flow which is the difference of the parameters describing the collective motion of free 
neutrons and protons weighted by their numbers~\cite{li02}. As pointed out 
by Yong et al. \cite{yong06}, this observable minimizes the influence of the isoscalar 
part in the EOS while maximizing that of the symmetry term. Its proportionality to the particle
multiplicities, however,  makes its determination very dependent on the experimental efficiencies of
particle detection and identification and on the precise theoretical distinction between free and
bound nucleons.  
In this work, the ratios of neutron versus proton or light-charged-particle flows will be 
considered and their sensitivities 
to the strength of the symmetry term will be explored.
The problem of systematic errors is thereby reduced.

In a series of experiments at the GSI laboratory combining the LAND and FOPI (Phase 1) detectors, 
both neutron and hydrogen collective flow observables from $^{197}$Au + $^{197}$Au 
collisions at 400, 600 and 800 MeV/nucleon have been measured~\cite{leif93,lamb94}. 
These data sets are 
presently being reanalyzed in order to determine optimum conditions 
for a dedicated new experiment, but also with the aim to produce constraints
for the symmetry energy by comparing with predictions of state-of-the-art 
transport models. Here, we report the results obtained with the 400-MeV/nucleon data set
and the Ultrarelativistic-Quantum-Molecular-Dynamics (UrQMD) model which 
has recently been adapted to heavy ion reactions at intermediate energies~\cite{qli05}.
 
\begin{figure}[tb]
\centering
\includegraphics*[width=80mm]{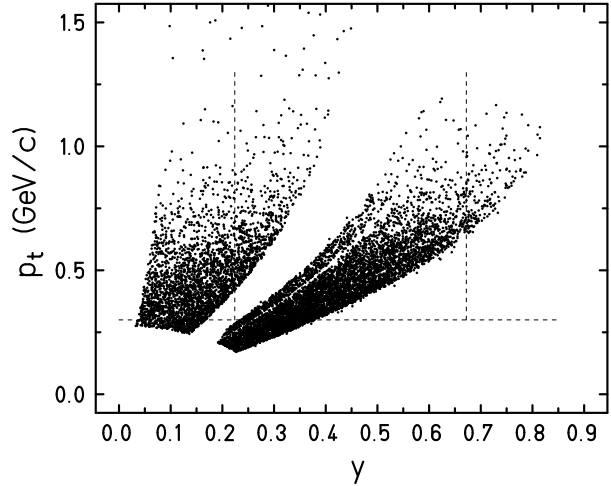}
\vskip -0.1cm
\caption{Scatter plot of 10000 neutron events with energy $E_{\rm lab} \ge 40$~MeV 
from a run without shadow bars in the plane of transverse momentum $p_t$ vs. rapidity $y$. 
The quality criteria are the same as used in the flow analysis. Their application produces
the local inefficiency in the forward detector unit caused by slabs with reduced performance. 
The dashed lines represent the acceptance cuts $p_t \ge 0.3$~GeV/c and $0.25 \le y/y_p \le 0.75$ 
applied later in the analysis.
}
\label{fig_0}
\end{figure}

The Large-Area-Neutron-Detector LAND is a 2x2x1 m$^3$ calorimeter consisting of in total 200 slabs
of interleaved iron and plastic strips viewed by photomultiplier tubes at both ends~\cite{LAND}.
For the flow experiment, LAND was divided into two parts of 50-cm depth each, positioned 
to cover laboratory angles of 37$^{\circ}$--53$^{\circ}$ and 61$^{\circ}$--85$^{\circ}$. 
The corresponding acceptance in the plane of transverse momentum $p_t$ vs. rapidity $y$ is
shown in Fig.~\ref{fig_0}.
Veto walls of 5-mm-thick plastic scintillators in front of the two detector units permitted 
the distinction of neutral and charged particles and to resolve atomic numbers $Z \le 2$.
A shower-recognition algorithm is used to identify individual particles within the registered
hit distributions.
Particle momenta were determined from the measured time-of-flight ($\Delta t \approx 
550$~ps, FWHM) over flight paths of 
5~m and 7~m to the more backward and more forward detector units, respectively. 
The amount of light collected from identified showers was used to generate mass spectra of the
hydrogen isotopes as described in Ref.~\cite{lamb94}. 
Individual isotopic resolution is not achieved but the strong proton group is clearly identified.
The selected proton samples, containing about 45\% of the total hydrogen yields, 
are expected to have isotopic purity of better than 95\%.

During its Phase 1, the 4-$\pi$ detector FOPI consisted of a forward plastic array, built 
from more than 700 individual plastic-scintillator strips and covering the forward hemisphere
at laboratory angles from 1$^{\circ}$ to 30$^{\circ}$~\cite{FOPI}. 
LAND was used to detect and identify neutrons and light 
charged particles with nearly identical methods while the FOPI Forward Wall was used to determine 
the modulus and orientation of the impact parameter from the multiplicity $M_c$ and 
distribution of the detected charged particles~\cite{reis97}. The background of scattered 
neutrons was measured in separate runs with shadow bars made
from iron and positioned to block the direct flight paths from the target to one or the other
of the two LAND units. Additional details can be found in Refs.~\cite{leif93,lamb94}.

The usefulness of the studied observables has been evaluated with the UrQMD transport 
model~\cite{qli05}. This model, originally developed to study particle production at high 
energy~\cite{bass98}, has been adapted to intermediate-energy heavy-ion collisions by introducing
a nuclear mean field corresponding to a soft EOS with momentum dependent forces and with 
different options for the dependence on asymmetry. Two of these choices are used here, 
expressed as a power-law dependence of the potential part of the symmetry energy on the
nuclear density $\rho$ according to
\begin{equation}
E_{\rm sym} = E_{\rm sym}^{\rm pot} + E_{\rm sym}^{\rm kin} 
= 22~{\rm MeV} \cdot (\rho /\rho_0)^{\gamma} + 12~{\rm MeV} \cdot (\rho /\rho_0)^{2/3}
\label{eq:pot_term}
\end{equation}
with $\gamma =0.5$ and $\gamma =1.5$ corresponding to a soft and a stiff density dependence 
($\rho_0$ is the normal nuclear density of $\approx 0.16$~fm$^{-3}$).
The kinetic part remains unchanged.
The UrQMD model has been successfully used for studies of a wide range of heavy-ion-collision
problems at intermediate and relativistic energies (Ref.~\cite{qli09} and references
given therein).
 
\begin{figure}[tb]
\centering
\includegraphics*[width=80mm]{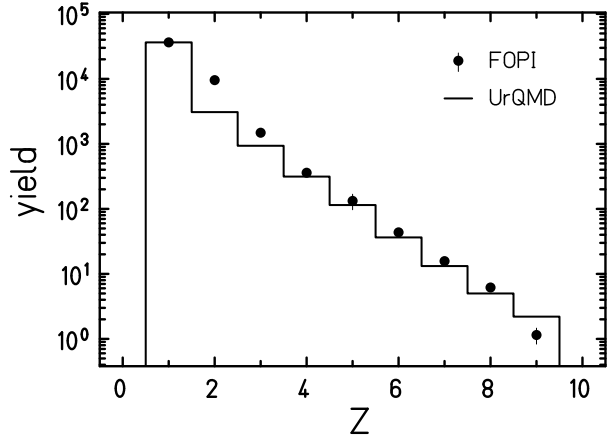}
\vskip -0.1cm
\caption{Fragment yields, integrated over the 4-$\pi$ solid angle, 
in central (equivalent to $b<2.0$ fm) collisions of
$^{197}$Au + $^{197}$Au at 400 MeV/nucleon as a function of $Z$ 
(dots, from Ref.~\protect\cite{reis97}) 
in comparison with UrQMD predictions normalized at $Z = 1$ (histogram).
}
\label{fig_1}
\end{figure}

As mentioned above, a realistic description of the clustering processes during the temporal 
evolution of the reaction is crucial for predicting dynamical properties of free neutrons, 
protons and light charged particles. In the UrQMD, the clustering algorithm is based on the
proximity of nucleons in phase space with two parameters for the relative nucleon coordinates 
and momenta. The present analysis was performed with the distributions obtained after 
a reaction time 
of 150~fm/c and with the proximity limits $\Delta r = 3.0$~fm and $\Delta p = 275$~MeV/c which
are typical for QMD models~\cite{zhang06}. These parameters had been slightly adjusted from 
their default values in order to obtain a best possible description of measured fragment 
spectra as a function of $Z$. 
The result for central collisions of $^{197}$Au + $^{197}$Au at 400 MeV/nucleon
is shown in Fig.~\ref{fig_1} in comparison with the data of Reisdorf et al.~\cite{reis97}.
With a normalization at $Z=1$, the overall dependence on $Z$ is rather well reproduced but
the yields of $Z=2$ particles are underpredicted by about a factor of 3. 
The strong binding of $\alpha$ particles is beyond the phase-space criterion used in the model.
However, also the 4-$\pi$ integrated yields of deuterons and tritons in central collisions 
are underestimated by similar factors of 2 to 3.

\begin{figure}[htb!]
\centering
\includegraphics*[width=80mm]{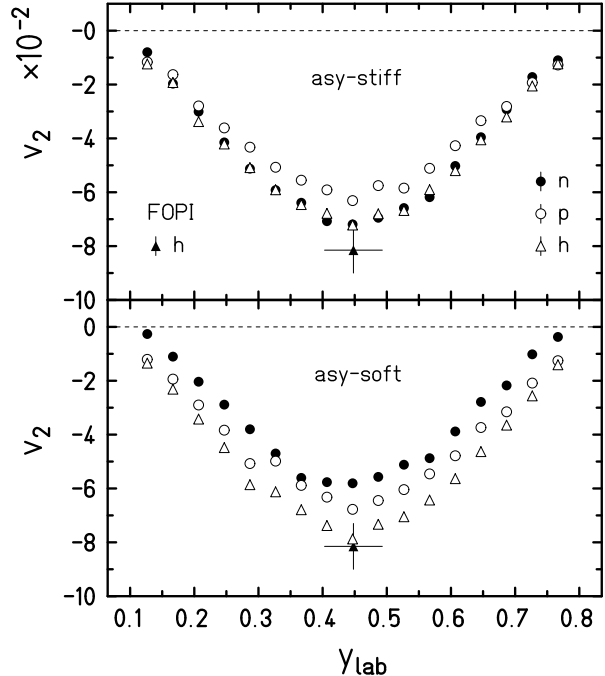}
\vskip -0.1cm
\caption{Elliptic flow parameter $v_2$ for mid-peripheral (5.5 $\le b \le$ 7.5 fm) 
$^{197}$Au + $^{197}$Au 
collisions at 400 MeV/nucleon as calculated with the UrQMD model for neutrons (dots), 
protons (circles), and all hydrogen isotopes ($Z=1$, open triangles), integrated over 
transverse momentum $p_t$, as a function of the laboratory rapidity $y_{\rm lab}$. 
The predictions obtained with a stiff and a soft density dependence of the symmetry term are 
given in the upper and lower panels, respectively. The experimental result from 
Ref.~\protect\cite{andro05} for $Z=1$ particles at 
mid-rapidity is represented by the filled triangle (the horizontal bar represents the experimental
rapidity interval).
}
\label{fig_2}
\end{figure}

Flow observables were obtained from the calculated event samples by fitting the azimuthal
particle distributions with the usual Fourier expansion 
\begin{equation}
f(\Delta\phi) \propto 1 + 2\cdot v_1\cdot {\rm cos}(\Delta\phi) +
2\cdot v_2\cdot {\rm cos}(2\cdot \Delta\phi)
\label{eq:fourier}
\end{equation}
with $\Delta\phi$ representing the azimuthal angle of the momentum vector of the 
emitted particle with respect to the reaction plane~\cite{andro06}. 
The predictions obtained for the elliptic flow of neutrons, protons, and 
hydrogen yields for $^{197}$Au + $^{197}$Au at 400 MeV/nucleon and for the two choices 
of the density dependence of the symmetry energy, 
labeled asy-stiff ($\gamma = 1.5$) and asy-soft ($\gamma = 0.5$),
are shown in Fig.~\ref{fig_2}.
The dominant difference is the significantly larger neutron squeeze-out in the asy-stiff case 
(upper panel) compared to the asy-soft case (lower panel). The proton and hydrogen 
flows respond only weakly, and in opposite direction, 
to the variation of $\gamma$ within the chosen interval. 

Collective flow has recently been measured rather systematically~\cite{andro05}. The
reliability of the methods has been demonstrated by the good agreement of data sets from
different experiments in the overlap regions of the studied intervals in collision 
energy~\cite{andro06}. An excitation function of flow for $^{197}$Au + $^{197}$Au 
collisions has been produced for incident energies from below 100 MeV/nucleon up to the 
ultrarelativistic regime.
Within this wide range, the negative elliptic flow of light particles, i.e. their squeeze-out 
perpendicular to the reaction plane, has its maximum at 400 MeV/nucleon. The value 
$v_2 = -0.082$ measured at this energy for $Z=1$ particles at mid-rapidity~\cite{andro05} 
is reasonably well reproduced by the UrQMD model, even though underestimated 
by 8\% and 15\% with the two parameterizations (Fig.~\ref{fig_2}).

The predictions for elliptic particle flows depend strongly on the type of transport 
theory used and on its parameters. 
Values lower than found experimentally have been reported also for other models as, e.g., 
discussed in Refs.~\cite{zhang06,giordano10}. In a recent paper, Li et al. 
investigate the elastic in-medium nucleon-nucleon cross section and show the effects of 
its momentum dependence on various observables~\cite{qli10}. The standard parameterization,
also used here up to now, is labeled FP1 in that paper. One of the alternatives, FP2, 
exhibits a higher correction factor than FP1 for relative nucleon-nucleon momenta 
around $0.2 < p_{\rm NN} < 0.5$~GeV/c at high density. With
this parameterization, the elliptic flow is larger, now exceeding the measured $Z=1$ value
by 16\% and 19\% for the stiff and soft parameterizations of the symmetry term, respectively. 

\begin{figure}[htb!]
\centering
\includegraphics*[width=80mm]{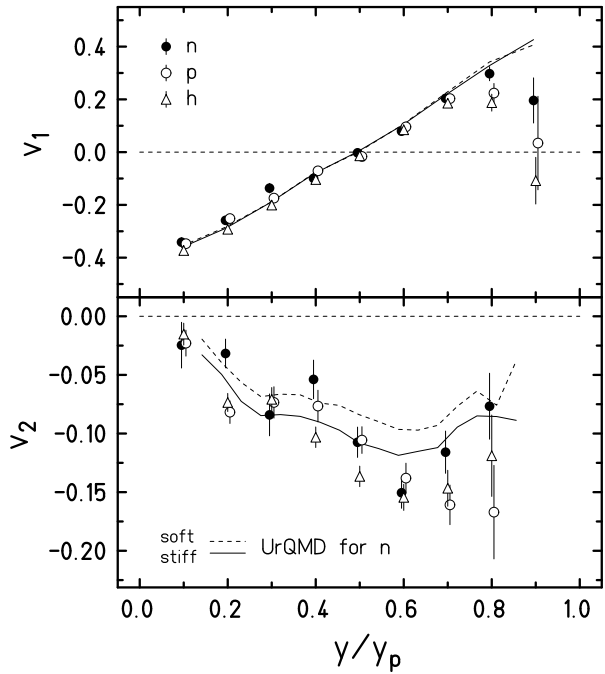}
\vskip -0.1cm
\caption{Measured flow parameters $v_1$ (top) and $v_2$ (bottom)
for mid-peripheral (5.5~$\le b \le$~7.5 fm) $^{197}$Au + $^{197}$Au collisions at 400 MeV/nucleon
for neutrons (dots), protons (circles), and hydrogen isotopes ($Z=1$, open triangles) 
integrated within $0.3 \le p_t/A \le 1.3$~GeV/c/nucleon
as a function of the normalized rapidity $y_{\rm lab}/y_p$. 
The UrQMD predictions for neutrons, obtained with a stiff ($\gamma = 1.5$, full lines) and a 
soft ($\gamma = 0.5$, dashed) density dependence of the symmetry term have been filtered to 
correspond to the geometrical acceptance of the experiment. The experimental data have been 
corrected for the dispersion of the reaction plane (see text).
}
\label{fig_3}
\end{figure}

The results obtained from the FOPI/LAND data set for 400 MeV/nucleon 
are shown in Fig.~\ref{fig_3} as a
function of the rapidity $y$, normalized with respect to the projectile rapidity $y_p = 0.896$,  
and for the PM3 event class ($27 \le M_c \le 39$) which approximately corresponds to the 
interval of impact parameters $5.5$~fm~$<b<7.5$~fm~\cite{reis97}. 
The geometrical acceptance of the LAND 
detector in this experiment produces a dependence of the accepted range in transverse momentum 
$p_t$ on rapidity $y$, $p_t$ increases with $y$ (Fig.~\ref{fig_0}), and is responsible for the 
observed asymmetry of $v_1$ and $v_2$ with respect to mid-rapidity $y/y_p=0.5$. Gates were set
for transverse momenta $0.3 \le p_t/A \le 1.3$~GeV/c, mainly to ensure equal lower thresholds for
neutrons and charged particles detected with LAND. For the comparison, the UrQMD outputs have been 
filtered with the geometrical acceptance of the LAND detector and subjected to the same 
transverse-momentum gates as the experimental data. 
The experimental Fourier coefficients, on the other hand, have been corrected for the 
uncertainty associated
with the reconstruction of the reaction plane from the azimuthal distribution of particles
measured with the FOPI Forward-Wall detector. Because of the significant multiplicities
$M > 26$ of particles observed in the considered range of impact parameters $b \le 7.5$~fm,
this correction is not crucial~\cite{andro06}, and global correction factors 1.05 for $v_1$
and 1.15 for $v_2$ have been adopted.

Only the 
predictions for neutrons, obtained with the standard FP1 parameterization of the elastic
nucleon-nucleon cross section~\cite{qli10},
are shown in Fig.~\ref{fig_3}, for clarity as well as because of 
the much smaller dependence on $\gamma$ of the results for hydrogen isotopes 
(cf. Fig.~\ref{fig_2}).
A nearly negligible sensitivity to the stiffness of the symmetry energy 
is exhibited by the directed flow, according to the UrQMD model (Fig.~\ref{fig_3}, top
panel). The lines representing the predictions for the soft and stiff parameterizations of 
the symmetry energy fall practically on top of each other. 
The results for protons or hydrogen isotopes are nearly identical.  
All predictions compare well, however, with the experimental results for the multiplicity 
bin PM3 corresponding to this range of mid-peripheral impact parameters. The range of rapidities
$y \ge 0.8$ at which the data deviate from the expected linearity coincides with the
shift of the LAND acceptance to transverse momenta $p_t \ge 0.75$~GeV/c/nucleon at which
the yields start to drop rapidly (Fig.~\ref{fig_0}).

Also the second Fourier coefficient $v_2$ describing elliptic flow and its dependence
on rapidity are fairly well described by the UrQMD calculations (Fig.~\ref{fig_3}, bottom panel). 
Here the statistical errors of the experimental data are larger but the slight underprediction
expected from Fig.~\ref{fig_2} is clearly recognized.  
Contrary to the directed flow, the $v_2$ values calculated with the asy-stiff and asy-soft 
parameterizations for neutrons are significantly different from each other. 
This is not the case for the hydrogen isotopes whose squeeze-out parameters are close to those
of neutrons for the asy-stiff case or, for free protons alone, similar to the asy-soft neutron
predictions.
Even though the experimental errors are considerable, the sensitivity of the neutron squeeze-out
seems large enough to permit the extraction of a useful result from the present data set.

\begin{figure}[htb!]
\centering
\includegraphics*[width=80mm]{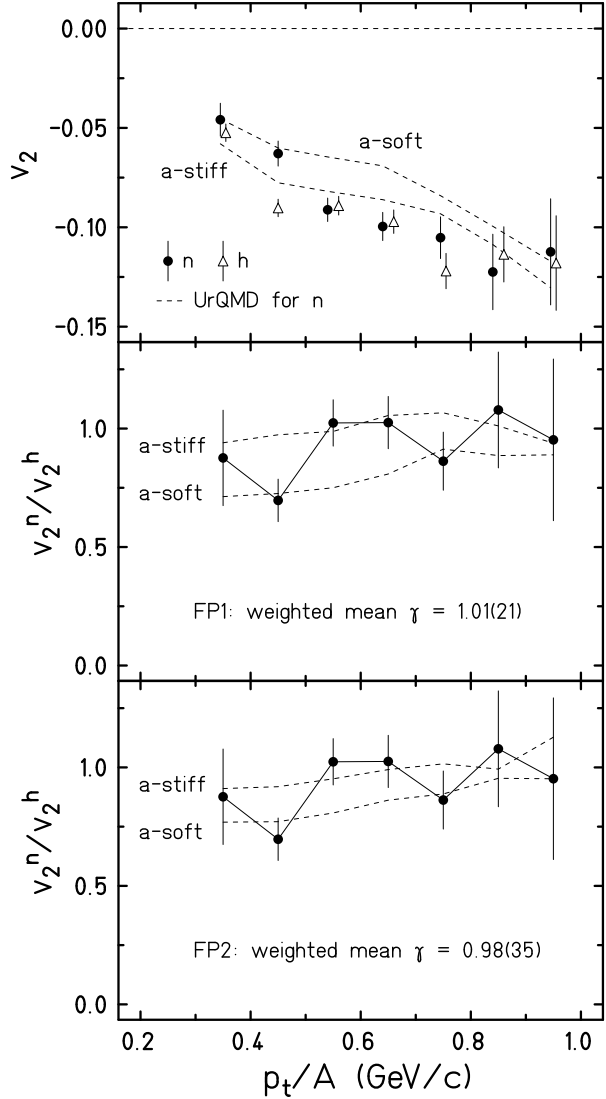}
\vskip -0.1cm
\caption{Differential elliptic flow parameters $v_2$ for neutrons (dots) and 
hydrogen isotopes (open triangles, top panel) and their ratio (lower panels) for moderately 
central ($b<7.5$ fm) collisions of $^{197}$Au + $^{197}$Au at 400 MeV/nucleon, integrated
within the rapidity interval $0.25 \le y/y_p \le 0.75$, as a function 
of the transverse momentum per nucleon $p_t/A$. The symbols represent the experimental data.
The UrQMD predictions for $\gamma = 1.5$ (a-stiff) and $\gamma = 0.5$ (a-soft)
obtained with the FP1 parameterization for neutrons (top panel) and for 
the ratio (middle panel), and with the FP2 parameterization for the ratio (bottom panel)
are given by the dashed lines.
Note the suppressed zero of the abscissa.
}
\label{fig_4}
\end{figure}

The dependence of the elliptic flow parameter $v_2$ on the transverse momentum per nucleon,
$p_t/A$, is shown in Fig.~\ref{fig_4}, upper panel, for the combined data set of central 
and mid-peripheral collisions (PM3--PM5, $M_c \ge 27$, cf. Ref.~\cite{reis97}) 
in order to exploit the
full statistics collected in the experiment. The extreme rapidities are avoided by
selecting the interval $0.25 \le y/y_p \le 0.75$. The increase of $v_2$ 
in absolute magnitude is nearly linear and the measured values for neutrons and hydrogen 
isotopes follow approximately the UrQMD predictions for $b<7.5$~fm. For hydrogens (not
shown), the results for the stiff and soft density dependences are similar and 
close to the asy-stiff predictions for neutrons. For neutrons, however, the predictions 
are different by, on average, 20\% for the main part of the $p_t$ interval (dashed lines) 
but seem to converge at high $p_t$ at which the yields become small and the statistical 
errors large.

For the quantitative evaluation, the ratio of the flow parameters of neutrons versus 
hydrogen isotopes is proposed to be used. Systematic effects influencing the
collective flows of neutrons and charged particles in similar ways will be minimized in this
way, on the experimental as well as on the theoretical side. Among them are, in particular, 
the still existing uncertainties of isoscalar type in the EOS but also, e.g., the dispersion
of the experimentally determined orientation of the reaction plane, possible small azimuthal 
anisotropies in the particle detection with the Forward Wall, and the matching of the
impact-parameter intervals used in the calculations with the experimental event groups selected
according to multiplicity. 

The results for the ratio $v_2^n/v_2^h$, i.e. with respect to the integrated hydrogen yield, 
are shown in Fig.~\ref{fig_4} (lower panels). They exhibit once
more the sensitivity of the elliptic flow to the stiffness of the symmetry energy predicted by 
the UrQMD. It is slightly smaller with the FP2 parameterization of the in-medium nucleon-nucleon 
cross section (bottom panel). The experimental ratios, even though associated with large errors, 
are found to scatter within the intervals given by the calculations for $\gamma = 0.5$ and 1.5
in either case, indicating that the effects of the cross section parameterization largely 
cancel in the ratios. Linear interpolations between the predictions, averaged over  
$0.3 < p_t/A \le 1.0$ GeV/c with weights $(\Delta \gamma_i)^{-2}$
derived from the experimental errors $\Delta \gamma_i$ for each $p_t$ bin,
yield $\gamma = 1.01 \pm 0.21$ and $\gamma = 0.98 \pm 0.35$ for 
the FP1 and FP2 parameterizations, respectively. 

The same analysis was also performed for the squeeze-out ratios $v_2^n/v_2^p$ of neutrons with 
respect to free protons.
The resulting power-law exponents are $\gamma = 0.99 \pm 0.28$ and $\gamma = 0.85 \pm 0.47$ for
the FP1 and FP2 parameterizations, respectively.
They are nearly identical to those obtained from the neutron-to-hydrogen squeeze-out ratios,
even though associated with larger statistical uncertainties. 
If the comparisons of the neutron-to-hydrogen ratios $v_2^n/v_2^h$ are 
restricted to the mid-peripheral interval of impact-parameters $5.5 \le b<7.5$~fm,
smaller values $\gamma = 0.58 \pm 0.27$ and $\gamma = 0.35 \pm 0.44$
are obtained with the FP1 and FP2 parameterizations, respectively. This apparent dependence on
impact parameter existed already in preliminary results reported earlier~\cite{traut09}. 

All results are, within errors, compatible with each other. In particular, the fact that 
the yields of deuterons and tritons with respect to protons are underestimated in the model 
calculations does not seem to have visible consequences. The ratios with respect to all
hydrogen isotopes and to protons alone yield the same exponent $\gamma$.
However, because of the limited statistical accuracy, the existence of systematic uncertainties
cannot be ruled out and may, in fact, be partly responsible for the apparent dependence on 
impact parameter.
We will, therefore, assume that the difference between the weighted means of
the four results for $b<7.5$~fm and the full statistics, $\gamma = 0.98$, and of the two values 
for $v_2^n/v_2^h$ and the mid-peripheral range $5.5 \le b<7.5$~fm, $\gamma = 0.52$, 
can serve as an estimate, $\Delta \gamma_{\rm |syst.} = \pm 0.23$, for the magnitude of 
possible systematic uncertainties.

The statistical errors of the elliptic flow coefficients translate into statistical errors of
$\gamma$ that depend on the sensitivity displayed by the predictions (Fig.~\ref{fig_4},
lower panels). Since the magnitude of elliptic flow is either slightly underpredicted or 
overpredicted with the two parameterizations, we interpolate between the two cases and obtain 
$\Delta \gamma_{\rm |stat.} = \pm 0.27$ as the most probable statistical error.
Combining it with the systematic uncertainty and considering the spread and trends of the 
individual results, we adopt the value $\gamma = 0.9 \pm 0.4$ as the result of this analysis 
for the exponent describing the density dependence of the potential term. 
Together with the kinetic term proportional to $(\rho/\rho_0)^{2/3}$, the squeeze-out data thus 
indicate a moderately soft behavior of the symmetry energy. 
It is obvious that data with higher statistical precision are highly desirable. 
Being able to follow the evolutions with impact parameter, transverse momentum 
(cf. Fig.~\ref{fig_4}) and particle type more accurately will provide supplementary constraints 
for the comparison with theory and will permit a clearer identification of systematic 
dependences related to it. 

Other systematic errors were found to be comparatively minor.
In the experimental data analysis, the explored variations of the conditions 
and gates within reasonable limits have only small effects. 
The uncertainty associated with the procedures used to measure and subtract the background of 
scattered neutrons detected with LAND~\cite{leif93,lamb94} has been estimated by subtracting 
only 60\% instead of 100\% of the measured background contributions from the total neutron 
yields. It reduces the resulting power-law exponent $\gamma$ by $|\Delta \gamma| \approx 0.2$ 
which may be considered an upper limit for systematic experimental effects. 

Additional systematic uncertainties on the theoretical side have also been evaluated. 
In particular, a test has been made whether supra-saturation densities are actually probed by
the squeeze-out ratios. The coefficient of the potential term in the symmetry energy has been
reduced from its default value 22 MeV to 18 MeV (cf. Eq.(\ref{eq:pot_term})). 
A softer (stiffer) density dependence is expected in this case if the symmetry term were 
mainly tested at densities below (above) saturation.
The result obtained for $v_2^n/v_2^h$ with FP1, $\gamma = 0.93 \pm 0.28$, 
is close to the corresponding $\gamma = 1.01 \pm 0.21$ obtained with 22 MeV, 
thus confirming that the dynamics at supra-saturation density are important for the comparison. 
The parameter $L$, proportional to the slope
of the symmetry term at saturation~\cite{lipr08} changes by less than 20 MeV when this 
modification is made. The adopted result $\gamma = 0.9 \pm 0.4$ corresponds to 
$L = 83 \pm 26$~MeV in the standard parameterization with 22 MeV.

In summary, the ratio of the elliptic-flow parameters of neutrons with respect to hydrogen isotopes 
measured for $^{197}$Au + $^{197}$Au collisions at 400 MeV/nucleon in the FOPI/LAND experiment 
has been used to determine the strength of the symmetry term in the equation of state 
by comparing it to UrQMD model predictions. A value of $\gamma = 0.9 \pm 0.4$ has
been obtained for the power-law coefficient describing the density dependence of the potential 
term. Individual systematic errors were not found to exceed a margin 
of $|\Delta\gamma| \approx 0.2$. Together with the kinetic term proportional to 
$(\rho/\rho_0)^{2/3}$, this indicates a moderately soft symmetry term. 
Within errors, it is consistent with the density dependence deduced from fragmentation experiments 
probing nuclear matter near or below saturation~\cite{klimk07,cent09,tsang09} but inconsistent 
with the super-soft or extremely stiff behavior obtained from the comparisons of the 
$\pi ^-/\pi ^+$ yield ratios measured for the same reaction with different 
transport models~\cite{xiao09,feng10}. Further experimental and theoretical work
will be needed to resolve these apparent ambiguities. 

Illuminating discussions with Lie-Wen~Chen, M.~Di~Toro, Bao-An~Li, W.~Reisdorf, and H.H.~Wolter 
are gratefully acknowledged. This work has been supported by the European Community under 
contract No. HPRI-CT-1999-00001 and FP7-227431 (HadronPhysics2), 
by the United Kingdom Science and Technology 
Facilities Council and Engineering and Physical Sciences Research Council, 
by the Ministry of Education of China under grant No. 209053, 
by the National Natural Science Foundation of China under grant Nos. 10905021 and 10979023, 
by the Zhejiang Provincial Natural Science Foundation of China under grant No. Y6090210, 
and by the Polish Ministry of Science and Higher Education under grant No. DPN/N108/GSI/2009.





\end{document}